\documentclass{article}
\pdfoutput=1
\usepackage[utf8]{inputenc}
\usepackage{amsmath}
\usepackage{amsfonts}
\usepackage{amssymb}
\usepackage{mathtools}
\usepackage{hyperref}
\usepackage{cleveref}
\usepackage{todonotes}
\usepackage{caption}
\usepackage[margin=0.5in]{geometry}
\usepackage{siunitx}
\usepackage{pythontex}
\usepackage{todonotes}
\usepackage{subcaption}
\usepackage{algorithm2e}
\usepackage{array,booktabs}
\usepackage{bbm}
\usepackage{rotating}
\usepackage{amsmath}
\usepackage[affil-it]{authblk}
\usepackage{hyperref}
\usepackage{xparse}

\newcommand\muw{\mu_{\text{Wilson}}}
\newcommand\muwhat{\hat{\mu}_{\text{Wilson}}}
\newcommand\sigwhat{\hat{\Sigma}_{\text{Wilson}}}
\newcommand\sigw{{\Sigma}_{\text{Wilson}}}
\newcommand\relion{diagonal}
\newcommand\rel{diagonal}
\newcommand\cryosparc{spatially-independent exponential}
\newcommand\cryos{SIE}

\newcommand\mur{\mu_{\text{\rel}}}
\newcommand\sigr{\Sigma_{\text{\rel}}}
\newcommand\PS{\text{PS}}

\newcommand\phiguess{\phi_{\text{guess}}}
\newcommand\hatphiguess{\hat{\phi}_{\text{guess}}}
\newcommand\Natoms{N_{\text{atoms}}}
\newcommand\vphi\varphi

\newcommand{\<}{\left\langle }
\let\>\undefined
\newcommand{ \>}{\right\rangle }
\newcommand{\mi}{\mathrm{i}} 
\DeclareMathOperator*{\argmax}{arg\,max}
\DeclareMathOperator*{\argmin}{arg\,min}
\DeclareMathOperator{\EX}{\mathbb{E}}

\newtheorem{assumption}{Assumption}

\title{ A Molecular Prior Distribution for Bayesian Inference Based on Wilson Statistics}
\author{Marc Aur\`ele Gilles\thanks{A.S. and M.A.G are supported in part by AFOSR FA9550-20-1-0266, the Simons Foundation Math+X
Investigator Award, NSF BIGDATA Award IIS-1837992, NSF DMS-2009753, and NIH/NIGMS
1R01GM136780-01}~\thanks{Program in Applied and Computational Mathematics, Princeton University, Fine Hall, Washington Road, Princeton, NJ 08544-1000, mg6942@princeton.edu}~, Amit Singer\thanks{Department of Mathematics and PACM, Princeton University, Fine Hall, Washington Road, Princeton, NJ 08544-1000}
 }
\date{\today}

\begin{document}

\maketitle
	
\begin{abstract}
\textbf{Background and Objective:}
Wilson statistics describe well the power spectrum of proteins at high frequencies. Therefore, it has found several applications in structural biology, e.g., it is the basis for sharpening steps used in cryogenic electron microscopy (cryo-EM). A recent paper gave the first rigorous proof of Wilson statistics based on a formalism of Wilson's original argument. This new analysis also leads to statistical estimates of the scattering potential of proteins that reveal a correlation between neighboring Fourier coefficients.
Here we exploit these estimates to craft a novel prior that can be used for Bayesian inference of molecular structures.

\textbf{Methods:}
We describe the properties of the prior and the computation of its hyperparameters. We then 
evaluate the prior on two synthetic linear inverse problems, and compare against a popular prior in cryo-EM reconstruction at a range of SNRs.

\textbf{Results:}
We show that the new prior effectively suppresses noise and fills-in low SNR regions in the spectral domain. Furthermore, it improves the resolution of estimates on the problems considered for a wide range of SNR and produces Fourier Shell Correlation curves that are insensitive to masking effects.

\textbf{Conclusions:}
We analyze the assumptions in the model, discuss relations to other regularization strategies, and postulate on potential implications for structure determination in cryo-EM.
\end{abstract}	
	

\section{Introduction}\label{sec:introduction}
In his seminal 1942 paper,
Arthur Wilson showed that the power spectrum of molecules is approximately flat at high frequencies~\cite{wilson1942determination}. 
The core assumption Wilson made is that atoms act as if they were randomly positioned at high frequency. Wilson used that randomness assumption and the observation that atom positions map to waves in the spectral domain to argue that the cross-terms of the waves sum up incoherently, implying a flat power spectrum.
Wilson statistics finds several applications in structural biology. For example, in the field of cryogenic electron microscopy (cryo-EM), it is the basis of B-factor sharpening~\cite{rosenthal2003optimal}, a post-processing step aimed at increasing the contrast of experimentally obtained reconstructions.

A recent paper~\cite{wilson_singer} provided a rigorous derivation of Wilson statistics based on a formalism of Wilson's argument.
Using that formalism, the author derived other forms of statistics, particularly a mean and covariance estimate for the scattering potential of molecules.
In the present paper, we 
develop a prior based on these estimates and study their use in Bayesian inference of protein structure in cryo-EM.
After we state background on the use of maximum a posteriori estimation and priors in cryo-EM in~\cref{sec:priors_cryo}, we describe the new prior in~\cref{sec:wilson_prior}. Then, in~\cref{sec:methods}, we evaluate the prior on synthetic linear inverse problems and discuss its properties and further applications in~\cref{sec:discussion}.

The code to reproduce numerical experiments appearing in
this manuscript is publicly available at:
\url{https://github.com/ma-gilles/wilson_prior}.

\section{Background} \label{sec:background}
\subsection{Priors in cryo-EM}\label{sec:priors_cryo}
In standard single particle cryo-EM structure determination,
 the 3D scattering potential map of a molecule $\phi:  \mathbb{R}^3 \rightarrow \mathbb{R} $ is determined from particle projection images $y_i:  \mathbb{R}^2 \rightarrow \mathbb{R}$. 
The critical computational step for reconstruction is the iterative refinement, typically formulated as maximum a posteriori estimation. In that step, the reconstructed molecule is the one that maximizes the posterior probability:
\begin{equation}\label{eq:MAP}
\phi_{\text{MAP}}  \coloneqq \argmax_{\phi} p( \phi | y ) \mathrel{\stackrel{{\mbox{\normalfont\tiny Bayes'  law}}}{=}}  \argmax_{\phi} p( y | \phi ) p(\phi) \ ,
\end{equation}
where $p( y | \phi)$ is the likelihood function (the conditional probability of the images given the molecule), and  $p(\phi) $ is the prior distribution over molecules. 
The prior encodes our belief about the distribution of molecules before any observation and often imposes particular properties on the inferred scattering potential (e.g., smoothness). 
Priors are a form of regularization: they add information to solve an ill-posed problem and avoid overfitting. 

In many inverse problems, including cryo-EM reconstructions, the quality of the inferred estimate depends heavily on the choice of prior. Thus, the cryo-EM community has given much attention to crafting appropriate priors.
One of the earliest software packages to use MAP estimation for 3D reconstruction in cryo-EM was RELION~\cite{scheres2012bayesian}.
The initial version of RELION used a Gaussian prior distribution where each frequency is independent with zero mean and variance equal to the spherically-averaged power spectrum of the molecule at that frequency. We call this prior the diagonal prior, denoted as:
\begin{equation}\label{eq:RELION_prior}
p_{\text{\rel}} ( \hat{\phi} ) := \mathcal{N}( 0, D_{\PS(\phi)}) \ ,
\end{equation}
where $\hat{\phi}$ denotes the Fourier transform of $\phi$, $\mathcal{N}(\mu, \Sigma)$ denotes a normal distribution with mean $\mu$ and covariance $\Sigma$, $D_{v}$ denotes a diagonal matrix with diagonal $v$, 
and $\PS(\phi) (\xi) : = \int_{S^2} |\hat{\phi} ( \|\xi\| \omega )|^2 d \omega $ denotes the spherically-averaged power spectrum of $\phi$. 
The \relion~prior has several attractive properties: it enforces smoothness on the reconstructed molecule, and it is computationally cheap thanks to the independence assumption of Fourier components.

The initial version of cryoSPARC~\cite{punjani2016building}, another popular software for cryo-EM reconstruction, used a \cryosparc~(\cryos) prior:
\begin{equation}\label{eq:cryoSPARC_prior}
p_{\text{\cryos}} ( \phi ) := \text{Exp}(\phi)
\end{equation}
where $\text{Exp}(\mu)$ denotes the exponential distribution with mean $\mu$. The advantages of this prior are that it imposes positivity on the scattering potential, and it is computationally convenient thanks to the independence of voxel values.
We note that despite the statistical interpretation of priors in~\cref{eq:MAP}, both the \relion~and \cryos~priors are chosen out of mathematical and computational convenience rather than a belief about the distribution of molecules. This may bias the reconstruction process~\cite{bendory2020single}.
In newer versions of RELION~and cryoSPARC, both software packages use a modified version of the \relion~prior as defined above. 
In that prior, each Fourier frequency is modeled as independent Gaussian with variance set proportionally to the SNR as estimated by Fourier Shell Correlation (FSC) between half maps~\cite{scheres2012RELION,punjani2017cryoSPARC}.

In recent years, \textit{implicit} regularization schemes have become popular in the computational imaging community, where there is often no explicit prior function $p(\phi)$. Instead, an operator mimics the 
regularizer's action in an iterative algorithm (e.g., the operator may act as a gradient descent step~\cite{romano2017little,reehorst2018regularization} or a proximal operator~\cite{venkatakrishnan2013plug,buzzard2018plug}).
One example in cryo-EM, implemented in RELION, uses a denoising neural network to regularize expectation-maximization iterations~\cite{kimanius2021exploiting}.
Another notable example, implemented in the non-uniform refinement option of cryoSPARC, regularizes the iteration using a smoothing kernel whose parameters are set adaptively by cross-validation~\cite{punjani2020non}.
Implicit regularization schemes are more general than regularization using an explicit prior\footnote{E.g., the impossibility result in~\cite{reehorst2018regularization}describes the conditions under which an implicit denoising regularizer cannot be written as an explicit regularizer.}  and can leverage powerful machine learning techniques
that sometimes yield impressive results, e.g.~\cite{kimanius2021exploiting, punjani2020non} both report that the prior nearly halves the attained resolution on some test cases compared to the traditional prior.
On the other hand, implicit regularization schemes lose most theoretical guarantees and statistical interpretation granted by Bayesian inference, sometimes leading to overfitting.
Overfitting is particularly problematic in cryo-EM, where ground truth about the 3D structure is often unavailable, making diagnosing overfitting difficult.
In the rest of this paper, we focus on building an explicit prior, similar to the ones in~\cref{eq:RELION_prior,eq:cryoSPARC_prior} but derived from Wilson statistics.
 
\subsection{The Wilson prior}\label{sec:wilson_prior}
\begin{figure}
\centering
\includegraphics[width=0.8\textwidth]{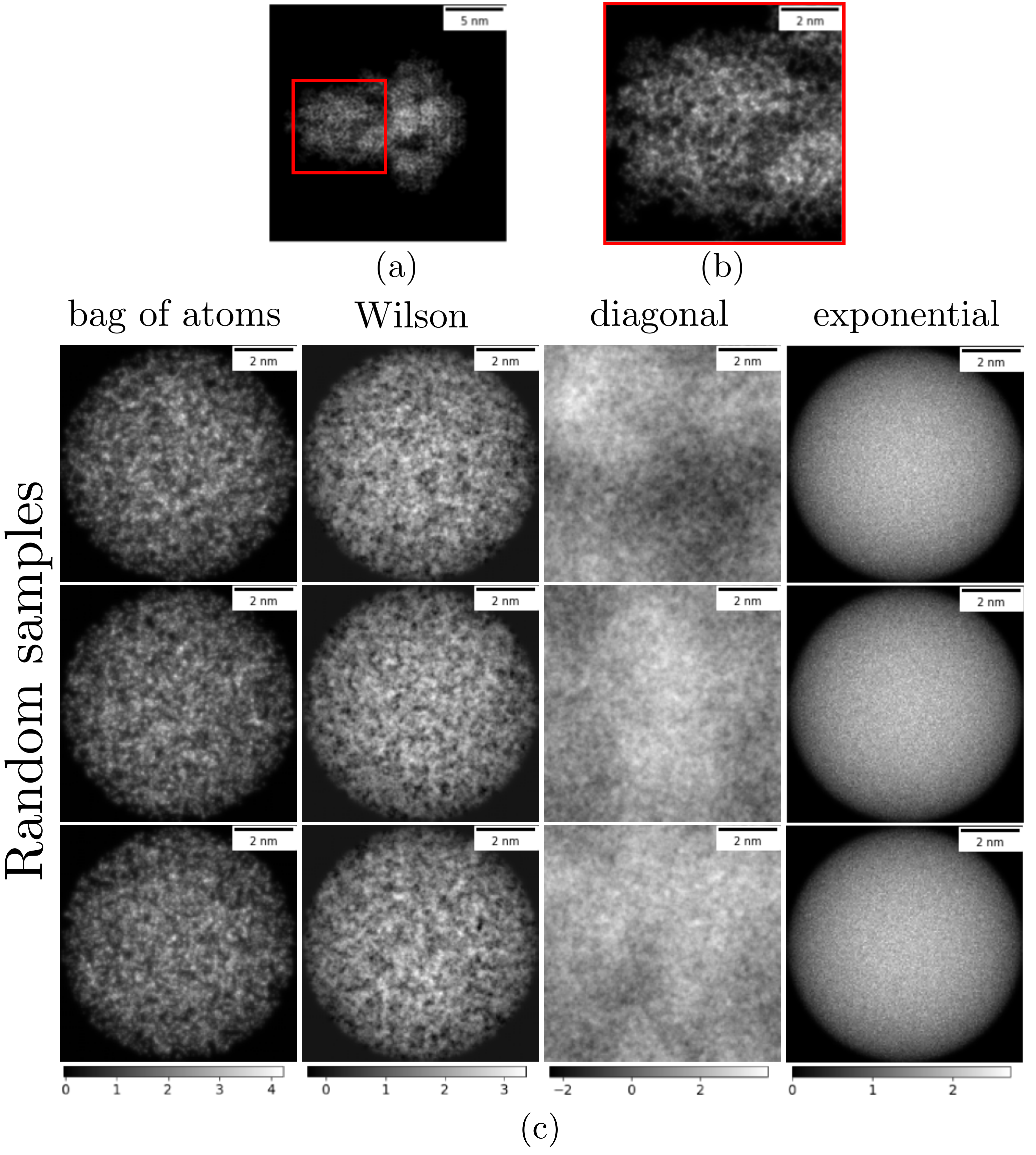}
    \caption{Illustration of different prior distributions and comparison with a reference protein. 
(a) Projection image of the SARS-CoV-2 spike glycoprotein (PDB 6VXX), used for reference. (b) Zoomed-in view of the spike protein at the same scale as images sampled from the priors; meant to illustrate the texture of real proteins at this scale.
(c) Random samples from four distributions: the bag of atoms model, the Wilson prior, the \relion~prior, and the \cryosparc~prior.
(1): the bag of atoms model with $\Natoms = 8871$, the shape function $g$ is a uniform distribution over the ball of radius 40 $\AA$ and the atom scattering function $ f$ is the carbon atom shape described in~\cite{peng1996robust}.
(2): the Wilson prior with the same $\Natoms$, $g$ and $f$.
(3): the \relion~prior where the variance of the frequencies is set to the expected power spectrum of the bag of atoms model.
(4): the \cryos~prior with mean equal to the expected value under the bag of atoms model.
The Wilson prior's distribution is close to the bag of atoms model but contains a small number of negative values. 
The \relion~prior has a similar texture but does not capture the support or voxel distribution. 
The \cryos~prior does not capture the texture of projection images due to the independence assumption of voxels in real space.
Mismatches between the prior and the actual distribution of molecules can bias the reconstruction by Bayesian inference. We emphasize that both the \relion~and \cryos~prior are chosen for computational convenience, whereas the Wilson prior was derived from a molecular model.
}    \label{fig:priors} 
\end{figure}

The formalism used to derive Wilson statistics in~\cite{wilson_singer} is the random ``bag of atoms" model.
In that model, a molecule consists of a collection of $\Natoms$ atoms whose positions $X_1, X_2, \dots, X_{\Natoms}$ are independently and identically distributed (i.i.d.) with probability density function $g: \mathbb{R}^3 \rightarrow \mathbb{R}^+$ that models the molecule's shape.
The scattering potential of a molecule $\phi: \mathbb{R}^3 \rightarrow \mathbb{R} $ is modeled as
\begin{equation}\label{eq:bag_of_atoms}
\phi ( x) = \sum_{i=1}^{\Natoms} f(x - X_i)
\end{equation}
where 
$f$ is the scattering potential of a single atom. For simplicity, here we assume that the atoms are identical and revisit to this assumption in~\cref{sec:discussion}.
If $f$, $g$, and $\Natoms$ are fixed,~\cref{eq:bag_of_atoms} implicitly defines a probability distribution over molecules that can, in principle, be used as a prior for Bayesian inference. Unfortunately, there is no easy formula for the probability density function of that distribution, so it is impractical to do so.
Instead, we
use the first two moments of the bag of atoms model to craft a Gaussian prior that resembles the original distribution but is easier to compute. 
We express this prior in the Fourier domain, but an equivalent formulation in the spatial domain is described in~\cref{sec:wilson_real}. 
The first two moments of the distribution of the Fourier transform of $\phi$ are derived in~\cite{wilson_singer}:
\begin{align}
\muwhat (\xi) & \coloneqq \EX [ \hat{\phi} (\xi) ] = \Natoms \hat{f}(\xi) \hat{g}(\xi)  \label{eq:wilson_mean} \ ,\\
\sigwhat (\xi_1, \xi_2) & \coloneqq
 \EX [ (\hat{\phi} (\xi) - \muwhat) \overline{(\hat{\phi} (\xi) - \muwhat)} ]= 
 \Natoms \hat{f}(\xi_1)  \overline{\hat{f}(\xi_2)} \left( \hat{g}( \xi_1 - \xi_2) -   \hat{g}(\xi_1)  \overline{\hat{g}(\xi_2) }\right)  \ . \label{eq:wilson_covariance}
\end{align}
where $\hat{f}$, $\hat{g}$ denote the Fourier transforms of $f$ and $g$ respectively.
We thus define the Wilson prior as:
\begin{equation} \label{eq:wilson_prior}
p_{\text{Wilson}}(\hat{\phi}) = N( \muwhat, \sigwhat  ) \ .
\end{equation}
Due to the non-Gaussianity of the bag of atoms model, the Wilson prior and the bag of atoms model are not the same. In particular, negative values in the spatial domain are possible in the Wilson prior but not in the bag of atoms model (assuming $f(x) \geq 0 $). Nevertheless, the Wilson prior is the 
maximum entropy distribution, which matches the first and second-order statistics derived from the bag of atoms model and allows us to derive a fast algorithm.
\Cref{fig:priors} illustrates the differences between the bag of atoms model and three priors: Wilson, \relion, and \cryos, as defined in~\cref{eq:wilson_prior,eq:RELION_prior,eq:cryoSPARC_prior}.

The \relion~and \cryos~priors represent two extremes: in the \relion~prior, frequencies in the spectral domain are independent, and in the \cryos~prior, voxels in the spatial domain are independent. The Wilson prior is a middle ground: the decay of the Fourier-transformed shape function $\hat{g}(\xi)$ dictates the correlation between two frequencies, and the decay of the function $f(x)$ dictates the correlation between two positions.
This latter fact is most evident from the Wilson statistics expressed in the spatial domain (see~\cref{sec:wilson_real} in the appendix).
In~\cref{fig:priors}, observe that projection images generated by the \relion~prior have a similar texture as the ones generated by the Wilson prior, but they do not capture the shape of the molecule.
In~\cref{sec:appendix}, we explain this phenomenon by showing that the \relion~and Wilson priors are equal in the limit on a uniform shape function on the entire domain.
As a result, we interpret the Wilson prior as a generalization of the \relion~prior beyond the case of a uniform shape function on the entire domain. 

If no information about the molecule's shape is available, 
an uninformative prior reflecting equal probability of the position of atoms at any point in the domain is appropriate. Thus, the \relion~prior is the logical choice.
However, during the iterative refinement of 3D reconstruction, we can often observe the low frequency of the shape function at early iterations.
Remarkably, these \textit{low frequencies} capture most of the \textit{correlation of high frequencies}. 
This is a consequence of a result in~\cite{wilson_singer}: under mild assumptions on the shape function, its Fourier transform decays quadratically with frequency $|g(\xi) | = \mathcal{O}(\xi^{-2}) $.
At high frequency, the decay implies that 
$|\hat{g}(\xi_1 - \xi_2)| \gg |\hat{g}(\xi_1) \hat{g}(\xi_2)|$ therefore the covariance of the Wilson statistics is: 

\[
 \sigwhat ( \xi_1, \xi_2 ) =  \Natoms \hat{f}(\xi_1) \overline{\hat{f}(\xi_2)} \left( \hat{g}(\xi_1 - \xi_2)   - \hat{g}(\xi_1) \hat{g}(\xi_2)\right) \approx \Natoms \hat{f}(\xi_1) \overline{\hat{f}(\xi_2)} \hat{g}(\xi_1 - \xi_2)  \ .
\]
Since $\hat{f}(\xi)$ decays slowly in the spectral domain, the quadratic decay of the shape function $\hat{g}(\xi)$ dictates the correlation
of neighboring frequencies.
Therefore, Fourier coefficients at high frequencies are correlated to their neighboring Fourier coefficients, but that correlation decays quadratically with the neighborhood size.
It follows that capturing $ \hat{g} (\xi ) $ for small $\xi$ (that is, the low frequencies of the shape function) is sufficient to capture most of the correlation between high frequencies. 
It is this correlation we seek to exploit in developing the Wilson prior.

\section{Application to inverse problems}\label{sec:methods}
\subsection{Problem statement}
We evaluate the Wilson prior on linear inverse problems; that is, by inferring the 3D scattering potential of the molecule $\phi$ from measurements $y$, generated by:
\begin{equation} \label{eq:linear_inv}
y = A \phi + \epsilon \ ,
\end{equation} 
where
$A$ is the measurement matrix, and $\epsilon \sim \mathcal{N}(0, \sigma^2 I)$ is additive Gaussian white noise. 
In particular, we study the denoising problem where $A$ is the identity and the deconvolution problem where $A$ is diagonal in the Fourier domain. 
Both problems are significantly simpler than 3D reconstruction of cryo-EM structures, but the expectation step of expectation-maximization takes the form of~\cref{eq:linear_inv}; where the diagonal matrix $A$ also accounts for marginalizing over the latent variables.

We estimate $\phi$ from $y$ with MAP estimation as in~\cref{eq:MAP}.
When the noise and the prior are Gaussian distributions (as in the case for Wilson and \relion~priors), the optimization problem is rewritten as a least-squares problem:
\begin{equation}\label{eq:MAP_log}
\phi^{MAP} = \argmin_{\phi} \left\lbrace \frac{1}{2\sigma^2 } \|  A \phi - y \|_2^2 + \frac{1}{2}\| \phi - \mu \|^2_{\Sigma^{-1}} \right\rbrace \ ,
 \end{equation}
where $\mu, \Sigma$ are the prior mean and prior covariance.
The solution of~\cref{eq:MAP_log} is the result of the Wiener filter:
\begin{equation}\label{eq:wiener_filter}
\phi^{MAP}  = ( I - HA) \mu + H y, \qquad  
H = \Sigma A^* ( A \Sigma A^* + \sigma^2 I)^{-1} \ .
\end{equation}
Plugging in different prior means and covariances result in different MAP estimates, which we refer to as MAP-\rel~and MAP-Wilson.
Both the \relion~and the Wilson priors have hyperparameters that depend on the true molecule: the spherically-averaged power spectrum in the case of the \relion~prior, the shape function $g$, atomic scattering function $f$, and the number of atoms in the case of Wilson prior.
In cryo-EM applications, the spherically-averaged power spectrum can often be well approximated from the data before the reconstruction is performed~\cite{singer2020computational}. 
However, the shape of the molecule is not typically known a priori. We address this in the following section.

\subsection{Estimating parameters of the Wilson prior} \label{sec:g_estimate}
The Wilson prior depends on three quantities: the number of atoms $\Natoms$, the atom scattering function $f(x)$ and the shape function $g(x)$.
To fix $\Natoms$ and $f(x)$, we assume that the atomic composition of the target molecule is known. We set $\Natoms$ to the ground truth number of non-hydrogen atoms and set $f$
 equal to the weighted average of the scattering functions of atoms in the molecule:
\[
f (x) = \sum_{a} p_a f_a(x)
\]  
where $a$ indexes the atom type, $p_a$ denotes the proportion of atom $a$ in the molecule, and $f_a$ the scattering potential of atom $a$, each modeled by a sum of 5 Gaussians as in~\cite{peng1996robust}.

To approximate the shape function $g$, we assume that we have a (possibly low resolution) estimate of the molecule $\phi_{\text{guess}}$.
We get a formula for $g$ by approximating the mean of the Wilson statistics as $\phi_{\text{guess}}$:
\begin{equation*}
\hatphiguess (\xi) \approx \muwhat (\xi) = \Natoms \hat{f}(\xi)  \hat{g} (\xi)
\end{equation*}
Solving for $g(x)$, we get: $g(x) = \mathcal{F}^{-1} \left\lbrace \frac{ \hatphiguess (\xi) }{  \Natoms \hat{f} (\xi) } \right\rbrace$, i.e., $g(x)$ is proportional to our guess image deconvolved by the atom scattering function.
If $\phiguess$ were precisely the true molecule scattering function and all atom scatterings were equal to $f(x)$, then the estimated $g(x)$ would be a sum of delta function at the true atomic positions.
To account for noise and modeling mismatch, 
we propose to relax this approximation by convolving this quantity with a Gaussian kernel 
 $ G_{\nu} ( x) =   \exp \left( - \|x\|^2 / (2 \nu^2) \right) / (2 \pi \nu^2)^{3/2}  $. 
Finally, since
$g(x)$ is a probability distribution of the position of atoms, it must lie in the probability simplex $\Delta =  \{ g | \int g(x) dx = 1,  g(x) \geq 0 \}$. 
Numerically, this condition ensures that the covariance matrix in~\cref{eq:wilson_covariance} is positive semi-definite. 
To enforce this constraint, we orthogonally project our guess onto $\Delta$ (denoted $\Pi_{\Delta}$). In summary, we estimate the shape function $g$ from a guess of the molecule $\phiguess$ by:
\begin{equation}\label{eq:g_estimate}
g_{\phiguess, \nu} (x) = \Pi_{\Delta} \left( G_{\nu} ( x) *  \mathcal{F}^{-1} \left\lbrace \frac{ \hatphiguess (\xi) }{  \Natoms \hat{f} (\xi) } \right\rbrace  \right) \ .
\end{equation}
In all numerical experiments, we take $\phiguess$ to be the MAP-\rel~estimate and set the kernel width $\nu=1\AA$. We discuss alternatives to estimate $g(x)$ in~\cref{sec:discussion}.

We mention that convolving an empirical distribution by a kernel is known as kernel density estimation; a non-parametric method typically used to estimate a random variable's probability density function based on a finite data sample (e.g.,~\cite{gramacki2018nonparametric}). Here, we use it in a different framework since we never directly observe samples of atomic locations. The effect of kernel density estimation is to smooth the distribution, which is interpreted as regularizing the estimated distribution.
This strategy also recalls the regularization scheme of~\cite{punjani2020non}, where a non-stationary Butterworth kernel is used to regularize in cryo-EM reconstruction.
A significant difference is that we use the kernel to set a hyperparameter of the prior, whereas~\cite{punjani2020non} uses the kernel directly to regularize the molecule.

\subsection{Evaluation of the Wilson prior on synthetic data}
We construct a ground truth scattering potential $\phi$ from the reported atomic coordinates of the spike protein (PDB 6VXX) by evaluating the following sum on a regular 3D frequency grid using the NUFFT~\cite{barnett2019parallel,barnett2021aliasing}:
\begin{equation}
\hat{\phi}(\xi) = \sum_{a} \hat{f}_a(\xi) \sum_{i_a} \exp(-2  \pi \mi \< \xi, x_{i_a}\>)
\end{equation}
where $a$ is the atom type, and $\hat{f}_a(\xi)$ is the atomic scattering model reported in~\cite{peng1996robust}.
The final ground truth is obtained by inverse discrete Fourier transform and masking using the ground truth mask reported below, which suppresses oscillations caused by truncation of the spectrum.
  
We report the Fourier Shell Correlation (FSC) between the ground truth and reconstructed molecules,
both with and without masks. Masked FSC is the standard method to estimate resolution in cryo-EM, 
but
using an overly tight mask will inflate the FSC, resulting in an over-optimistic resolution estimate. 
On the other hand, using no masks often underestimates the map's resolution, as the noise in the background dominates the FSC at higher frequencies. We use the resolution criterion ``masked FSC equal to $0.5$" since we compare our reconstruction to the ground truth~\cite{rosenthal2003optimal}.
We use a ground truth mask generated by the software package EMDA~\cite{warshamanage2021emda} by placing a sphere of radius $3\AA$ at each atomic location, followed by dilation and convolution with a Gaussian.  

\begin{figure}[h!]

\includegraphics[width=\textwidth]{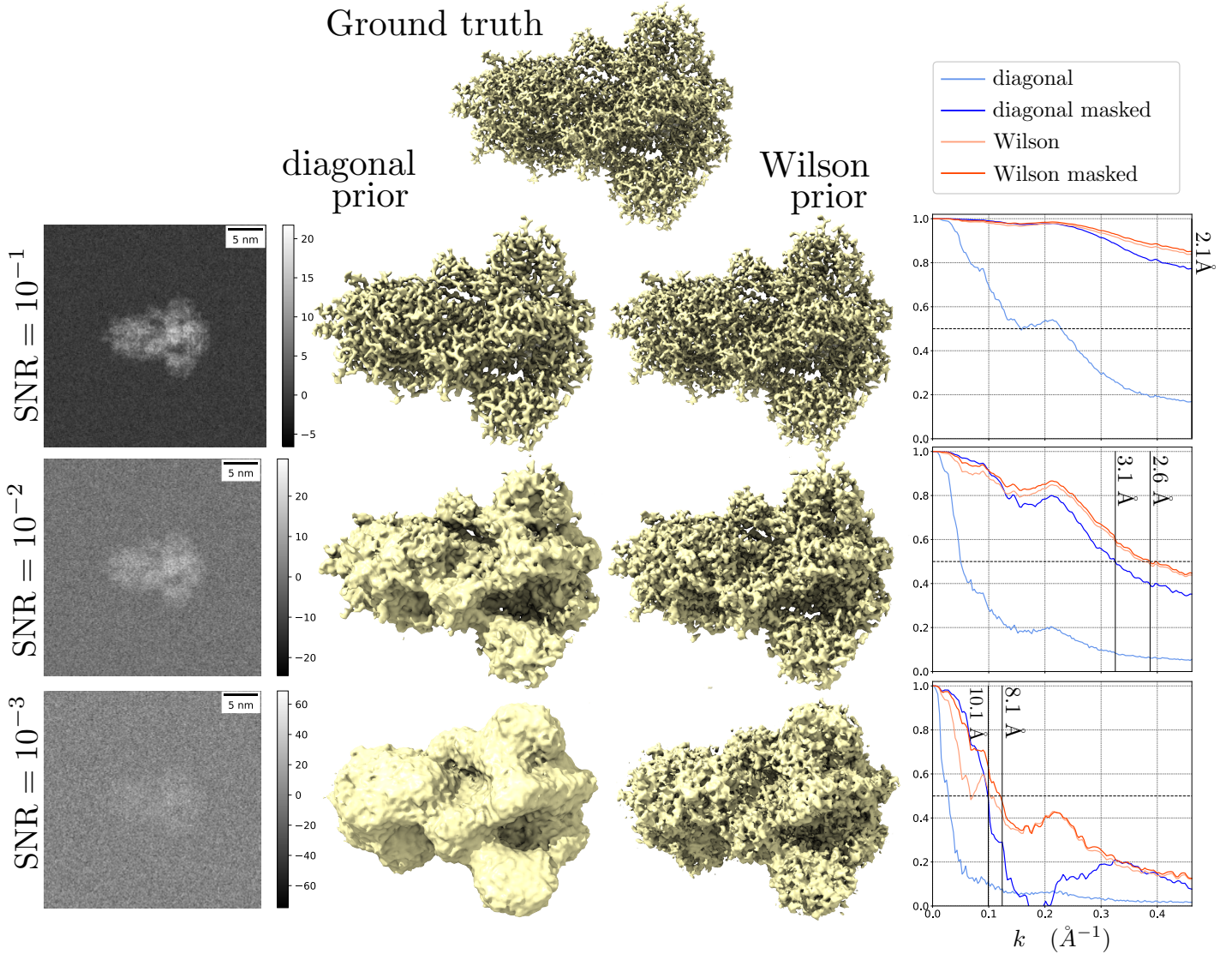}

\caption{ Denoising of the spike protein contaminated with Gaussian white noise at different SNRs. Left: projected images of the noisy observations. Middle: isosurface visualization of the ground truth, the MAP-diagonal estimate, and the MAP-Wilson estimate. Isosurface levels were qualitatively set to reflect the structure best. Right: FSC plots. 
\label{fig:spike_denoising}
}
\end{figure}

In~\cref{fig:spike_denoising}, 
we compare the result of the denoising Wiener filter using the~\relion~and Wilson prior at different SNRs.  
We define SNR $=\frac{\| \phi \|^2_2}{\sigma^2 N } $ where $N=301^3$ is the number of voxels.  
Following the implementation in~\cite{scheres2012bayesian}, we scale the covariance matrix of the diagonal prior (but not the Wilson prior) by a constant $T=4$. This scaling factor improves the reconstruction of the MAP-\rel~estimate, as was also empirically observed in~\cite{scheres2012bayesian}.
The first noteworthy result is that the FSC of the MAP-Wilson estimate is insensitive to the mask compared to the MAP-\rel~estimate. 
In~\cref{sec:wilson_real} in the appendix, we show that if we take the atomic scattering function to be a delta function (which is approximately true for low-resolution problems), the support of the shape function contains the support of the MAP-Wilson estimate.
Hence, estimating $g$ is analogous to calculating a mask, and the MAP-Wilson estimate has a masking effect.

Within the mask, the MAP-Wilson's resolution is significantly
higher than the MAP-\rel: $2.6 \AA$ vs $3.1\AA$ SNR = $10^{-2}$, and  $8.1 \AA$ vs $10.1 \AA$ at SNR = $10^{-3}$.
We interpret that the \relion~prior over-smoothes the protein, whereas the Wilson prior exploits\
the correlation of neighboring Fourier components
to suppress more noise while retaining higher frequency information.
Note that at SNR = $10^{-3} $, the resolutions are very low for both estimates ($10.1 \AA$ for \relion~vs. $8.1 \AA$ for Wilson), but a notable difference is the MAP-Wilson estimate is visually not smooth, unlike the MAP-\rel~estimate.
In this case, the smoothing of the \relion~prior might be a desirable property. We come back to this point in~\cref{sec:discussion}.

\begin{figure}[h!]

\includegraphics[width=\textwidth]{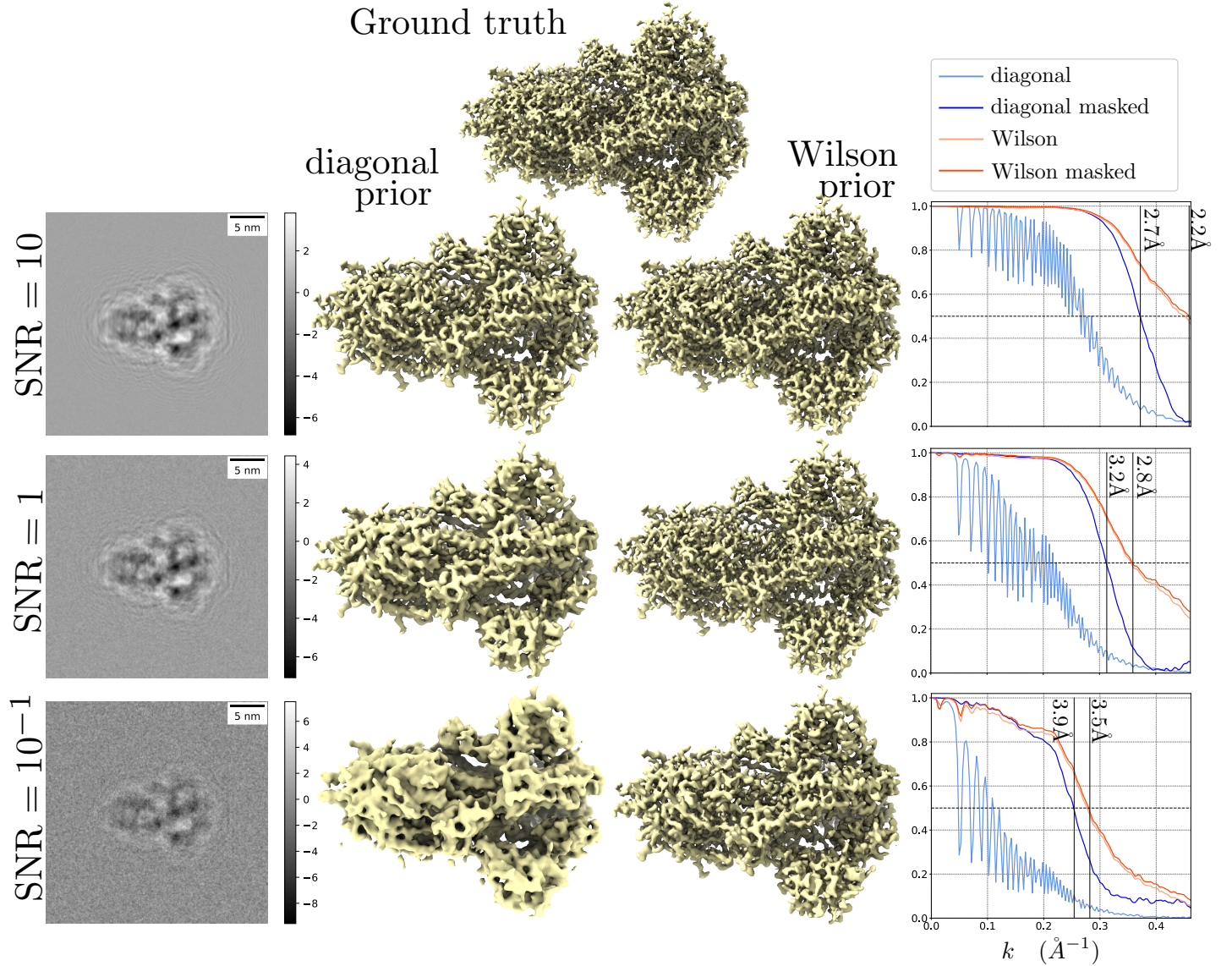}

\caption{Deconvolution of the spike protein contaminated with Gaussian white noise at different SNRs.
Left: projected images of the noisy observations.
Middle: isosurface visualization of the ground truth, MAP-diagonal estimate, and MAP-Wilson estimate.
Right: FSC plots.
\label{fig:spike_deconvolution}
}

\end{figure}

In~\cref{fig:spike_deconvolution}, we consider the problem of deconvolution with the contrast transfer function (CTF):
\begin{equation*}\label{eq:CTF}
\mbox{CTF}( \xi ) = \sin(- \pi \lambda d  \| \xi\| ^2 +  C_s \lambda^3  \| \xi\|^4 \pi/2  - \alpha)  \exp(- B  \| \xi\|^2 / 4) \ ,
\end{equation*}
with $\lambda = $ \SI{2.51}{\pico\metre}, $d = $ \SI{1.5}{\micro\metre}, $C_s = $\SI{2.0}{\micro\metre}, $\alpha = 0.1$, and $B = 80 \AA^2 $.
Note that the zero crossings of the CTF annihilate some of the measured Fourier coefficients; thus, the SNR is small around these frequencies. This produces the visible oscillations in the unmasked FSC of the MAP-\rel~estimate. In contrast, the oscillations are not present in the MAP-Wilson case. Therefore, we interpret that the Wilson prior uses the correlation between neighboring frequencies to fill zero-crossing gaps.
The ability to fill low-SNR gaps may be helpful in cryo-EM applications as SNR is non-uniform in Fourier space due to the CTF and the non-uniform distribution of molecule rotations. 

Similar to the denoising case, 
the FSC of the MAP-Wilson is  insensitive to masking, and
the resolution of the 
MAP-Wilson estimate is higher than the MAP-\rel~across all SNRs: $2.2 \AA$ vs $2.7\AA$ at SNR =$10$, $2.8\AA$ vs $3.2\AA$ at SNR = $1$ and $3.5\AA$ vs $3.9\AA$ at SNR = $0.1 $. 

\subsection{Computational matters}

The Wiener filter in~\cref{eq:wiener_filter} involves computing the result of 
\begin{equation}\label{eq:wiener_comp}
H v  = \Sigma A^* ( A \Sigma A^* + \sigma^2 I)^{-1} v \ .
\end{equation}
$\sigr$ and $A$ are diagonal matrices in the Fourier basis, therefore the MAP-\rel~estimate can be computed by elementwise multiplication in $\mathcal{O}(N)$ operations, where $N$ is the number of voxels.
However,
$\sigw$ is dense, so the same strategy does not apply. Instead, we solve the linear system in~\cref{eq:wiener_comp} with the conjugate-gradient algorithm (CG)~\cite{hestenes1952methods}. The computational bottleneck of CG is the matrix-vector products with the matrix  $ A \Sigma A^* + \sigma^2 I$. 
Thanks to the special structure of $\sigw$
 \footnote{$\Sigma$ is a  diagonally scaled convolution operator plus a rank one matrix. Diagonal scaling costs $\mathcal{O}(N)$ operations, convolution $\mathcal{O}(N \log (N))$ and the matrix vector product with a rank-1 matrix is $\mathcal{O}(N)$, thus the total cost is $\mathcal{O}(N \log (N))$.}
  these matrix-vector products can be computed in $\mathcal{O}(N  \log (N))$ using the FFT. 
The computational cost of solving~\cref{eq:wiener_comp} approximately is thus $\mathcal{O} (\kappa N \log (N))$ where $\kappa$ is the number of CG iterations. 
 In experiments, we terminate CG when $ \kappa = 100$ or we achieve a tolerance of $ 10^{-8}$.
Finally, computation of the shape function as described in~\cref{eq:g_estimate} costs $\mathcal{O} ( N \log ( N))$ operations with the FFT and fast projection algorithms for the probability simplex~\cite{wang2013projection}.

We note that although the \relion~prior is faster than the Wilson prior\footnote{The convergence of CG, and therefore the runtime of the Wilson prior, varies with the condition number of the matrix $(A\Sigma A^* + \sigma^2I)$, which depends on the CTF and the noise level.  
As an example, in our implementation, the denoising Wiener filter takes 26 seconds using the \relion~prior and 6 minutes using the Wilson prior for SNR=$10^{-2}$ and a grid of size $301^3$, on a Macbook Pro (16 GB RAM and 2 GHz Quad-Core Intel Core i5).
}, 
solving~\cref{eq:wiener_comp} for either prior is a negligible portion of the computation performed in 3D refinement.
Indeed, the dominant cost of the maximization step of a typical expectation-maximization iteration used in 3D refinement is marginalizing over latent variables and summing over projections images to form the matrix and the right-hand side, which dwarfs the cost of the CG iteration. The other additional computation is updating the prior mean and covariance (which one could consider part of the E step in an EM algorithm). Since we do not explicitly form the matrix $\sigw$, this only requires updating the shape function $g$. In this paper, we computed $g$ as in eq.~(11) at negligible computational expense, but more involved strategies such as the ones discussed in~section~4 could incur a more substantial cost. We leave this question to future work. 

\section{Discussion} \label{sec:discussion}

We showed a prior based on Wilson statistics outperforms a diagonal prior at solving two linear inverse problems within the MAP estimation framework.
We argue that the Wilson prior encodes more information about molecules, namely, the correlation between neighboring Fourier coefficients.
This correlation is used to suppress noise and fill gaps in low SNR regions of the spectral domain.

In contrast with implicit regularization, whose assumptions and convergence properties are opaque, we derived the Wilson prior from first principles. 
It fits squarely within the MAP framework, and we can guarantee its convergence and analyze violated assumptions.
The Wilson prior is the best Gaussian approximation of the bag of atoms model, which makes the following assumptions:
\begin{itemize}
\item \begin{assumption} \label{ass:f}
Atoms all have the same scattering function.
\end{assumption}
\item \begin{assumption} \label{ass:iid}
Atoms positions are independently identically distributed.
\end{assumption}
\item \begin{assumption} \label{ass:g}
Atoms positions are distributed with probability density function $g(x)$.\end{assumption}
\end{itemize}

We made~\cref{ass:f} for convenience here, but it is not necessary, and one could extend this framework to multiple atom types. 
However, numerical experiments indicate that~\cref{ass:f} is unimportant as there is a minimal loss of accuracy in denoising real molecules compared to fake molecules modified to have a single atom type.

~\Cref{ass:iid} is the core assumption made in Wilson statistics, but it is
violated in practice. For example, the position of subsequent atoms is fixed in an alpha-helix. 
Avoiding an independence assumption is difficult when crafting a prior, but considering atom interactions might bring further improvement if feasible. 

~\Cref{ass:g} can be problematic as a poor choice of $g(x)$ will lead to bias. For example, we showed that $g(x)$ acts as a mask in the denoising case, so carefully choosing its support is crucial.
We have advocated estimating $g(x)$ from an estimate of the molecule, but this could lead to overfitting, for the same reason that crafting a mask from an estimate may lead to overfitting~\cite{singer2020computational}.
On the other hand, we argued in~\cref{sec:wilson_prior} that only the lowest frequencies of $g(x)$ need to be accurately computed for most of the prior correlation to be correctly estimated. We also showed in synthetic experiments that a simple choice of $g(x)$ leads to significant resolution improvement with no visible overfitting. 
It is unclear whether this result will hold when the prior is used in an expectation-maximization algorithm.
In that case, a more robust framework may be to estimate $g$ as part of the maximum likelihood estimation.
The maximum likelihood estimate of $g$ and $\phi$ given measurements $Y$ is obtained by the following constrained minimization problem: 
\begin{align}\label{eq:MLE_with_g}
\phi_{\mbox{MLE}}, g_{\mbox{MLE}} = \argmin_{ g \in \Delta, \phi } \left\lbrace \frac{1}{2 \sigma^2}\| Y - A \phi \|_2^2 + \frac{1}{2} \| \Natoms \hat{f} \hat{g} - \phi \|^2_{ \Sigma_{g}^{-1}} + \log \det(\Sigma_{g}) \right\rbrace \ , \\
\text{where }\Sigma_{g}(\xi_1, \xi_2)  =  \Natoms \hat{f}(\xi_1) \left( g(\xi_1 - \xi_2) - \hat{g}(\xi_1) \overline{\hat{g} (\xi_2)} \right) \overline{\hat{f}(\xi_2) } \ .
\end{align}
Minimizing this quantity with an alternating scheme (e.g., ADMM~\cite{bertsekas2014constrained}) would produce iterations similar to ones outlined above: first minimize over $\phi$; which is equivalent to a Wiener filter as in~\cref{eq:wiener_filter}, and then minimize over $g$; corresponding to a non-linear version of the projection presented in~\cref{sec:g_estimate}.
Our results suggest that replacing mask estimation (a routine part of 3D reconstruction) with the estimation of the shape function would be beneficial for several reasons:
\begin{enumerate}
\item It improves the resolution of the estimates.
\item The corresponding FSC curves are insensitive to ground truth masks.
\item One can estimate the shape function within the Bayesian framework, which is typically not the case for a mask.
\end{enumerate}

To be integrated within a pipeline for 3D refinement,
the prior requires two generalizations: a variable noise model and an adaptive variance estimation. 
The former is a straightforward extension of the framework presented, e.g., the noise model used in~RELION can be implemented in~\cref{eq:wiener_filter} by replacing the matrix $\sigma^2 I$ by a different diagonal matrix.
The latter is a crucial property of the default prior used in RELION and cryoSPARC: the diagonal of the prior covariance matrix is updated at each iteration with the SNR as estimated by the FSC between half maps.
This strategy is a type of implicit frequency marching~\cite{bendory2020single}: the
regularization decreases for the high frequencies at each iteration, gradually increasing the resolution of the estimate. 
The same behavior does not exist for the prior we presented here since the variance of $\sigw$ at high frequencies is approximately $\Natoms | \hat{f}(\xi)|^2$, which is independent of the current estimate. 
One way to achieve the same result would be to scale the covariance matrix with the factor computed by FSC.
An alternative would be to treat the refinement as a blind deconvolution problem with a convolution kernel of the form $\exp(- \nu \|\xi\|^2) $.
If $\nu$ were updated using the FSC at each iteration, this step would be equivalent to B-factor correction, where Wilson statistics first found applications in cryo-EM. This may suggest that B-factor correction should be performed within each iteration rather than post-processing. 

\section{Conclusion}

We presented a prior distribution based on a simple yet expressive generative model for molecules. Then, we gave strategies to evaluate its hyperparameters and showed that, on simple inverse problems, the prior outperforms a prior similar to one often used in practice.
Finally, we discussed properties of the Wilson prior, its connections to other regularization strategies, and stated potential implications for structure determination in cryo-EM.

\section{Appendix} \label{sec:appendix}
\subsection{The Wilson prior in the spatial domain}\label{sec:wilson_real}
We establish the representation of the covariance operator $\sigw$ in the spatial domain:
\begin{align} \label{eq:wilson_cov_real}
\left( \sigw \phi \right) (x)  & = \Natoms f (x) * \left[  g(x) \left( \overline{f(x)}  * \phi (x) \right) \right] -    \Natoms f(x) *  g(x) \< g (x),  \overline{f (x)} * \phi(x) \>
\end{align}
where $*$ denotes convolution and $ \< \cdot, \cdot \>$ is the standard inner product on $L_2 (\mathbb{R}^3 )$. We
use the representation in the spectral domain established in~\cite{wilson_singer},
\begin{equation} \label{eq:wiener_cov_int}
(\sigwhat \hat{\phi}_2)(\xi_1) = \int_{\xi_2}  \sigwhat (\xi_1, \xi_2)  \hat{\phi}_2(\xi_2) d\xi_2 = \int_{\xi_2}\Natoms \hat{f}(\xi_1)  \overline{\hat{f}(\xi_2)} \left( \hat{g}( \xi_1 - \xi_2) -   \hat{g}(\xi_1)  \overline{\hat{g}(\xi_2) }\right) \hat{\phi}_2 (\xi_2) d\xi_2 \ .
\end{equation}

Let $\phi_1(x),\phi_2(x) \in L^2 (\mathbb{R}^3)$, and $\mathcal{F}$ 
denote the three-dimensional Fourier transform, then:
\begin{align}
\< \phi_1 (x) , \sigw \phi_2(x) \> 
&= \< \mathcal{F}^{-1} \mathcal{F} \phi_1(x), \sigw \mathcal{F}^{-1} \mathcal{F} \phi_2(x) \> \nonumber \\ 
&= \<  \mathcal{F} \phi_1(x), \left( \mathcal{F} \sigw \mathcal{F}^{-1} \right) \mathcal{F} \phi_2(x) \>\label{eq:line2}\\
&= \< \hat{\phi}_1(\xi_1),  \sigwhat \hat{\phi}_2(\xi_1) \> \label{eq:line2b} \\
&= \Natoms \<   \overline{\hat{f}(\xi_1)} \hat{\phi}_1(\xi_1), \int_{\xi_2 }  \overline{\hat{f}(\xi_2)} \left( \hat{g}( \xi_1 - \xi_2) -   \hat{g}(\xi_1)  \overline{\hat{g}(\xi_2) }\right) \hat{\phi}_2 (\xi_2) d\xi_2 \>\label{eq:line3} \\
&  \begin{aligned} =
 & \Natoms \<  \overline{\hat{f}(\xi_1)} \hat{\phi}_1(\xi_1), \int_{\xi_2 }    \hat{g}( \xi_1 - \xi_2)   \overline{\hat{f}(\xi_2)} \hat{\phi}_2(\xi_2) d\xi_2 \> \\ 
 &  \qquad - \Natoms \<   \overline{\hat{f}(\xi_1)} \hat{\phi}_1(\xi_1) , \hat{g}(\xi_1) \> \<  \hat{g}(\xi_2)  ,  \overline{\hat{f}(\xi_2)} \hat{\phi}_2(\xi_2) \> \nonumber
\end{aligned} \\
&  \begin{aligned} = &
\Natoms \<  \phi_1(x)  , f(x) * \left( g(x)  \left( \overline{f(x)} * \phi_2(x)  \right)  \right) \> \\
& \qquad -   \Natoms \<  \phi_1(x)  ,  f(x)  *  g(x) \> \<  g(x)  ,\overline{f(x)} * \phi_2(x)   \>  \label{eq:line5} 
\end{aligned} 
\end{align}
\Cref{eq:line2} follows from the unitary property of the Fourier transform $\mathcal{F}^* = \mathcal{F}^{-1}$,~\cref{eq:line2b} follows from the identity $\text{Cov}(A \phi) = A \text{Cov}(\phi)A^*$, \cref{eq:line3} follows from~\cref{eq:wiener_cov_int}, and~\cref{eq:line5} follows from the convolution theorem.
Since $\phi_1, \phi_2 \in L_2 (\mathbb{R}^3 )$ are arbitrary, this establishes~\cref{eq:wilson_cov_real}.
In quasi-matrix notation, the covariance operator is written as:
\begin{align*}
\sigw & = \Natoms  C_{f} \left( D_g - g g^* \right) C_{f}^*  
\end{align*}
where $C_f$ denotes the convolution operator with $f$, $D_g$ is the multiplication operator with $g$, and the outer product is defined as $(uv^*) \phi = u \<v, \phi \> $.
In the denoising problem $(A= I)$ and with the choice of $f(x)= \delta(x)$, which implies $C_f = I$, the Wiener filter matrix in~\cref{eq:wiener_filter} expressed in real space is:
\begin{equation*}
 H_{\text{Wilson}} = \Natoms \left( D_g - g g^* \right) ( \sigw + \sigma^2I)^{-1}
\end{equation*}
Note that in the case, $g(x)$ acts as a mask for the MAP-Wilson estimate:
\begin{equation*}
\phi^{MAP}_{\text{Wilson}}  = \muw + H_{\text{Wilson}} (y - \muw) = g \cdot w, \text{ where } w = \Natoms\mathbbm{1} +  \Natoms(I - \mathbbm{1} g^*) ( \sigw + \sigma^2I)^{-1} ( y - \muw) \ ,
\end{equation*}
where $\mathbbm{1}(x) = 1$.
That is, the MAP-Wilson estimate factorizes into to some function $w$ multiplied by the shape function $g$, which implies the support of $g$ contains the support of $\phi^{MAP}_{\text{Wilson}}$.

\subsection{The \relion~prior as the uniform limit of the Wilson prior}\label{sec:wilson_relion}
Recall that the Wilson and \relion~prior are defined as follows:
\begin{align*}
\mur (\xi) & =  0 \ , \\
\sigr (\xi_1, \xi_2 ) &= \begin{cases}
        \mbox{PS}(\| \xi_1\|) & \text{if } \xi_1 = \xi_2, \\
        0  & \text{else \ , } 
        \end{cases} \\
\muwhat (\xi) &=  \Natoms \hat{f}(\xi) \hat{g}(\xi)   \ ,\\
\sigwhat (\xi_1, \xi_2) &= 
 \Natoms \hat{f}(\xi_1)  \overline{\hat{f}(\xi_2)} \left( \hat{g}( \xi_1 - \xi_2) -   \hat{g}(\xi_1)  \overline{\hat{g}(\xi_2) }\right)  \ . 
\end{align*}
The diagonal of the \relion~covariance prior is the spherically-averaged power spectrum of the molecule.
Under the bag of atoms model, the expected power spectrum is (see~\cite{wilson_singer} for derivation):
\begin{align*}
\mbox{PS}(k) &= E \left[ \frac{1}{4 \pi} \int_{S^2} |\hat{\phi}(k \omega ) |^2  d \omega  \right] \\ & = |\hat{f}(k) |^2 \left( \Natoms+ \Natoms(\Natoms-1) \frac{1}{4\pi} \int_{S^2} |\hat{g}(k \omega)|^2 d\omega \right)  \\
\end{align*}
where we have assumed that $f(x)$ is a spherically symmetric function.
We now make the dependence on $\hat{g}$ explicit with superscripts, and even though that choice is not properly defined, we proceed with $\hat{g} = \delta$.
In that case, the mean and covariances of Wilson and \relion~agree for $\xi \neq 0$:
\[
\muwhat^{\hat{g} = \delta} (\xi) =  \Natoms \hat{f}(\xi) \delta(\xi)  =0 = \mur^{\hat{g} =\delta}(\xi)  \ ,\\
\]
and similarly for $\xi_1, \xi_2 \neq 0$: 
\begin{align*}
\sigwhat^{\hat{g} =\delta} ( \xi_1, \xi_2 ) &  =
\begin{cases}
  \Natoms  |\hat{f}(\|\xi_1\|) |^2 & \text{if } \xi_1 = \xi_2 \ , \\
        0  & \text{else } \ ,
        \end{cases}  \\ 
 \sigr^{\hat{g} =\delta} ( \xi_1, \xi_1 ) & = |\hat{f}(\|\xi_1\|) |^2 \left( \Natoms+ \Natoms(\Natoms-1) \frac{1}{4\pi} \int_{S^2} |\delta (k \omega)|^2 d\omega \right) =  \Natoms |\hat{f}(\|\xi_1\|) |^2  \ ,
\end{align*}
thus $\sigwhat^{\delta} (\xi_1, \xi_2) = \sigr^{\delta}(\xi_1, \xi_2) $. $\delta$ is not a valid choice of $\hat{g}$ as there is no well-defined probability density function $g$ for which $\mathcal{F} (g) = \delta(x)$.
Instead, we consider uniform distribution on a ball of radius $R$: $g_R(x) =\frac{3}{4\pi R^3} \mathcal{X}_B (x/R)$ where $\mathcal{X}_B$ is the characteristic function of the unit ball, and let the radius grow to infinity. The Fourier transform of $g_R$ is:
\[
\hat{g}_R(\xi) = - \frac{ 3 \cos(2 \pi R |\xi|)}{4 \pi^2 R^2 |\xi|^2} + \frac{3 \sin(2\pi R |\xi|)}{ 8 \pi^3 R^3 |\xi|^3}
\]
and we have the desired limit property $\lim_{R \rightarrow \infty} \hat{g}_R(x) = \delta(x)$.
Consequently:
\begin{align*}
\lim_{R \rightarrow \infty} \muwhat^{\hat{g} =\hat{g}_R}(\xi) = \lim_{R \rightarrow \infty} \mur^{\hat{g} =\hat{g}_R} (\xi) & \text{\qquad for } \xi \neq 0 \ , \\
\lim_{R \rightarrow \infty} \sigwhat^{\hat{g} =\hat{g}_R}(\xi_1, \xi_2) = \lim_{R \rightarrow \infty} \sigr^{\hat{g} =\hat{g}_R} (\xi_1, \xi_2) & \text{\qquad for } \xi_1, \xi_2 \neq 0 \ .
\end{align*}
That is, the \relion~and Wilson priors agree in the limit of a uniform distribution on the entire domain at non-zero frequencies.

\section*{Acknowledgement}
M.A.G. and A.S. are supported in part by AFOSR FA9550-20-1-0266, the Simons Foundation
Math+X Investigator Award, NSF BIGDATA Award IIS-1837992, NSF DMS-2009753, and
NIH/NIGMS 1R01GM136780-01. We thank Eric J. Verbeke for
valuable discussion and help in generating figures.

\bibliographystyle{spmpsci}
\bibliography{reference.bib}

\end{document}